\documentstyle[prl,aps,epsf,preprint]{revtex}
\begin{document}
\draft

\title{First principles study of Li intercalated carbon nanotube ropes}

\author{Jijun Zhao $^a$$^*$, Alper Buldum $^a$, Jie Han $^b$, Jian Ping Lu
$^a$ $^{\dagger}$}

\address{$^a$: Department of Physics and Astronomy, University of North
Carolina at Chapel Hill, Chapel Hill, NC 27599 \\
$^b$: NASA Ames Research Center, Mail Stop T27A-1, Moffett Field, CA 94035}

\date{\today}
\maketitle

\begin{abstract}

We studied Li intercalated carbon nanotube ropes by using first principles
methods. Our results show charge transfer between Li and C and small structural
deformation of nanotube due to intercalation. Both inside of nanotube and the
interstitial space are susceptible for intercalation. The Li intercalation
potential of SWNT rope is comparable to that of graphite and almost
independent of Li density up to around LiC$_2$, as observed in recent
experiments. This density is significantly higher than that of Li intercalated
graphite, making nanorope to be a promising candidate for anode material
in battery applications.

\end{abstract}

\pacs{71.20.Tx, 61.48.+c, 71.15.Pd, 68.65.+g}

Carbon nanotubes are currently attracting interest as constituents of novel
nanoscale materials and device applications \cite{1,2,3}. Novel mechanic,
electronic, magnetic \cite{2} and chemical properties \cite{3}  in these
one-dimensional materials have been discovered. Single-walled nanotubes
(SWNTs) form nanorope bundles with close-packed two-dimensional triangular
lattices \cite{4}. These rope crystallites might offer an all-carbon host
lattice for intercalation and energy storage. On analogy of the Li intercalated
graphite \cite{5}, carbon nanorope is expected to be a candidate of anode
materials for Li-ion battery applications \cite{6}. Recent experiments found 
much higher Li capacity (Li$_{1.6}$C$_6$) in SWNTs than those of graphite (LiC$_6$)
\cite{7}. The Li capacity can be further improved up to Li$_{2.7}$C$_6$ after
ballmilling the nanotube samples \cite{8}. This high capacity of Li in nanorope
implies lower weight and longer life time in the battery applications
\cite{9}.

First principles calculations have been successfully used to identify the 
cathode materials for  lithium batteries \cite{10}.
In previous theoretical works, K-doped small individual carbon nanotubes was
studied by first principles electronic structure calculations \cite{11}.
Empirical  force field model was also employed to simulate K doped SWNT ropes
\cite{12}.  However, there is no first
principles study on the Li-intercalated SWNT ropes. There are lots of open
questions such as: (1) what is the maximum Li intercalation
density; (2) where the intercalated Li ions sit; (3) what is the nature of
interaction between Li and the carbon nanotube; (4) does the intercalation 
modify the structure of nanotube. In this letter, we present results
obtained by first principles SCF pseudopotential calculations. Several
model systems of intercalated nanotube bundle are studied and the results
are discussed with the available experiments.

In this work, first principles SCF pseudopotential total-energy calculation and 
structural minimization are carried out within the framework of local-density 
approximation on a plane-wave basis with an energy cutoff of 35 Ry. The
Car-Parrinello algorithm with $\Gamma$ point approximation is used in the
electronic structure minimization \cite{13,14}. The ion-electron interaction
is modeled by Troullier-Martin norm-conserving nonlocal pseudopotential
\cite{15} in Kleinman-Bylander form \cite{16}. Plane-wave pseudopotential
program, CASTEP \cite{17}, is used for structural minimization on some 
selected systems.

\begin{figure}
\centerline{
\epsfxsize=3.0in \epsfbox{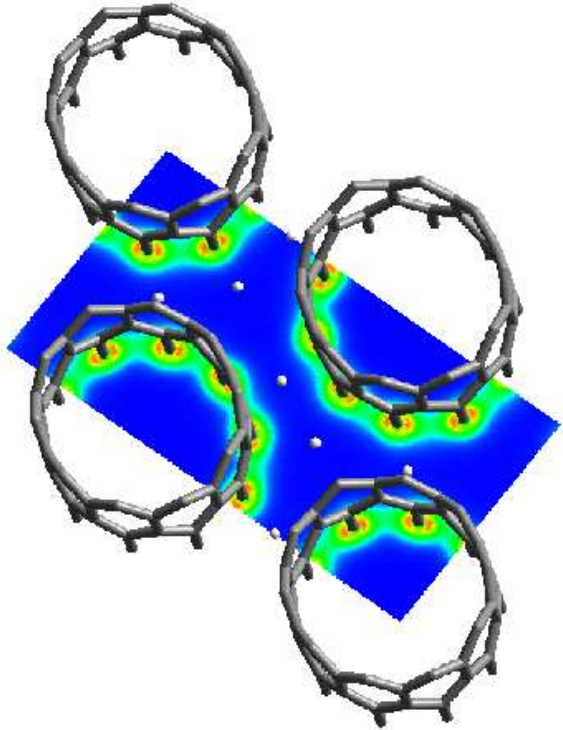}
}
\caption{Geometric structure and total electron density distribution (slice at 
(100) direction) of relaxed Li intercalated (10,0) tube bundle Li$_5$C$_{40}$. 
Small white balls denote Li atoms. Red, yellow, green, blue colors on the slice 
indicate electron density from higher and lower. No significant charge density on 
lithium sites is found, indicating charge transfer between Li and nanotubes.}
\end{figure}

The tube bundle is modeled by an uniform two dimensional hexagonal lattice.
The SWNTs studied here include both (10,0), (12,0) zigzag and armchair (8,8),
(10,10) tubes. The Li intercalated graphite and bulk Li are also
investigated as reference. The initial configuration of Li atoms are assumed 
to be on high-symmetric sites which maximize the Li-Li distance 
(see Table II for details).

In Fig.1, we plot the relaxed structure and charge density of (10,0) tube 
bundle with 5 Li atoms per unit cell. After structural minimization, the Li
atoms only slightly shift from their initial symmetric configuration.
This shows that symmetry and maximum Li-Li distance are good criteria for
choosing Li configuration. The Li intercalation also slightly modifies the
shape of carbon nanotubes (see Fig.1 and Fig.2). This result differs from
a previous empirical force field simulation on K doped (10,10) SWNT, in which
significant deformation on nanotubes was found \cite{12} (Using the same
force field model, we also obtain large deformation). The discrepancy between
first principle and empirical calculations demonstrates the importance of
quantum effect and the insufficiency of empirical potential in such systems.

\begin{figure}
\centerline{
\epsfxsize=3.0in \epsfbox{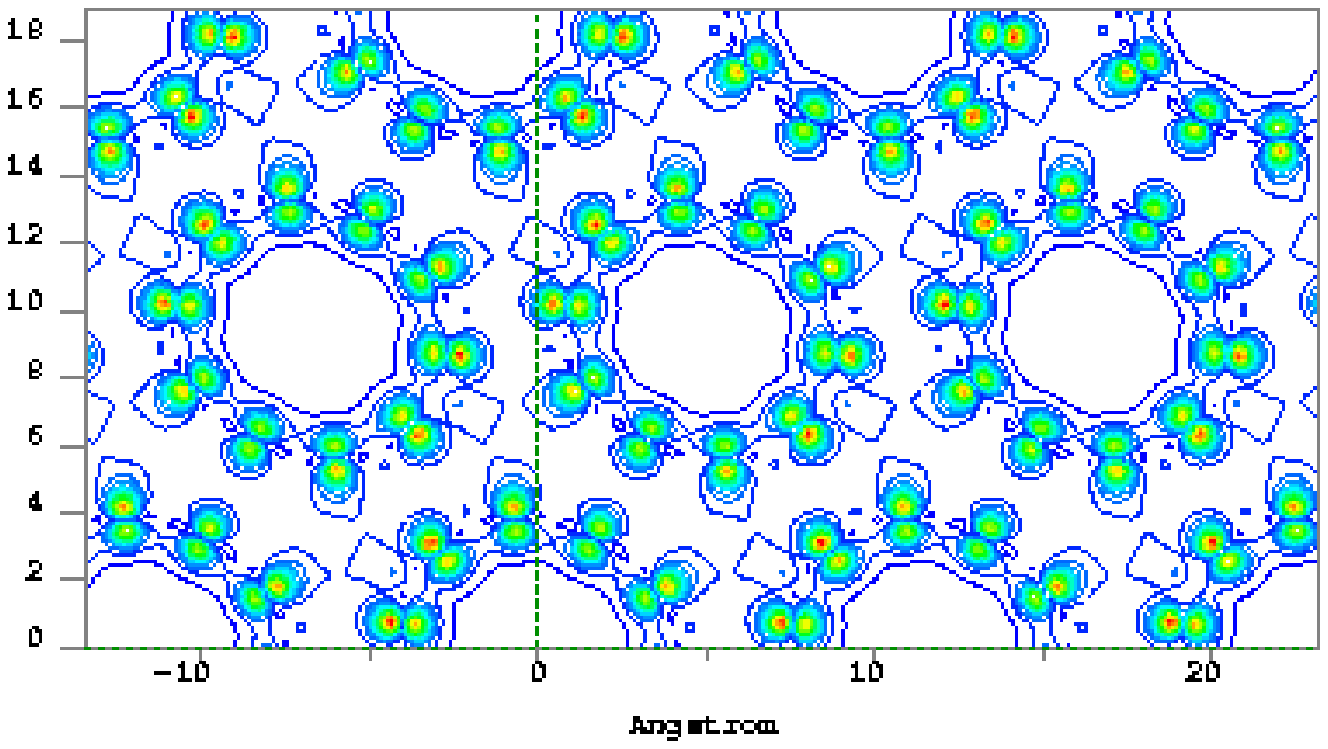}
}
(a)
\centerline{
\epsfxsize=3.0in \epsfbox{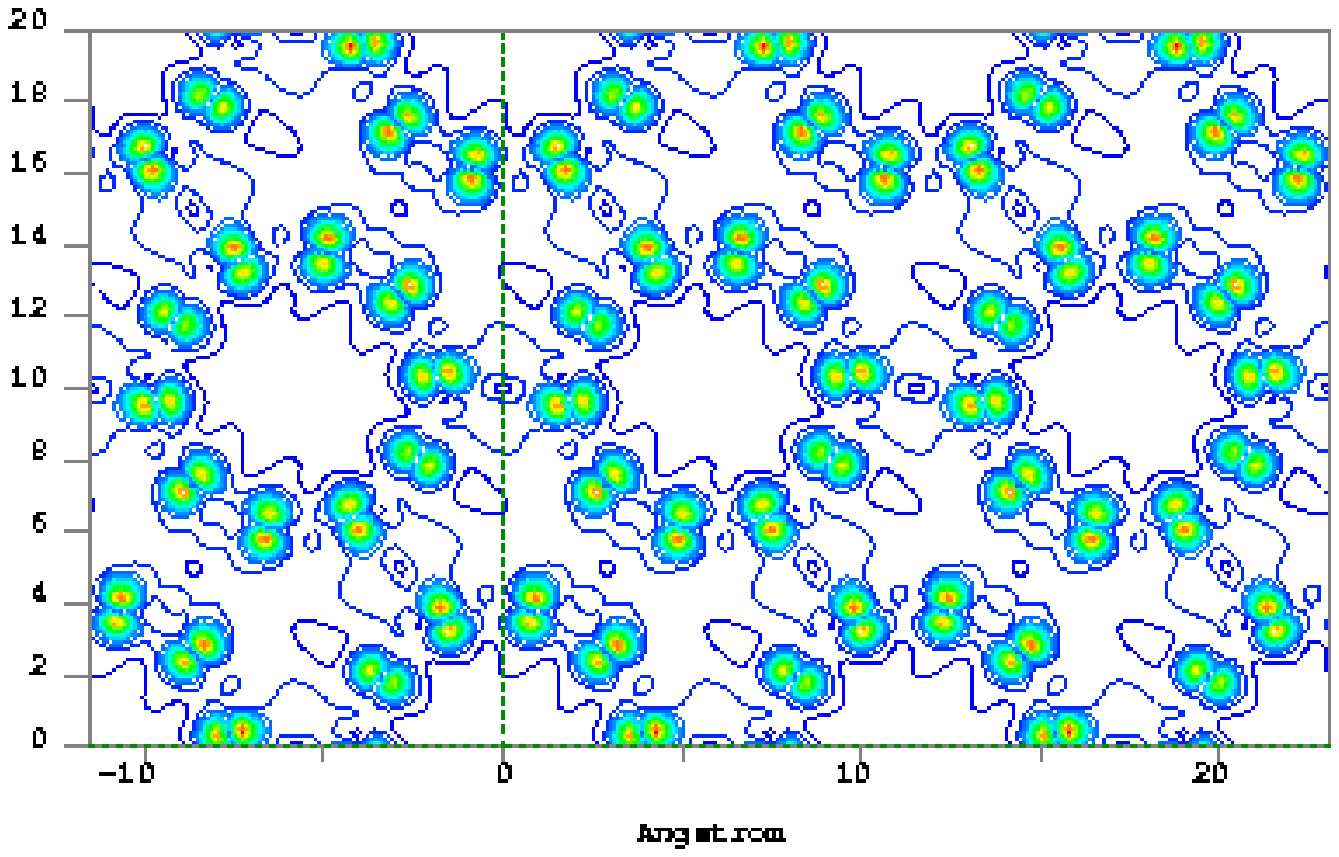}
}
(b)
\caption{Contour plots of the occupied conduction orbital densities near
Fermi level (a), and empty orbital densities above Fermi level (b) of
(10,0) tube bundle on the (001) plane. Red, yellow, green, blue colors
indicate electron density from higher and lower. Conduction orbitals are obviously
derived from $\pi$ bonds between C atoms. In (a), there is very low
conduction electron density pass through Li sites; whereas more distinct
distributions of empty states are found around Li ions in (b). Both indicate
charge transfer between Li and C.}
\end{figure}

Fig.1 shows that there is almost no total charge density distribution in the
space between SWNTs. To further understand the charge distribution, in Fig.2
we present the contour plots of both occupied (a) and empty (b) orbitals near
Fermi level. We find the conduction band orbitals are concentrated on the
carbon tubes, while the empty states has some distribution passing through
the Li sites.
In Fig.3, we compare the band structure near Fermi energy of tube bundle
and that of intercalated bundle. Although the individual (10,0) tube and
its bundle are all semiconductors, the intercalated tube bundle is
found to be metallic. For valence band, only small modification upon
intercalation is observed. In contrast, the hybridization between lithium
and carbon has significant influence on conduction band and introduces some
new states, similar to that found in Ref.[10]. All these analysis show that
there is almost complete charge transfer and the conduction electrons
mainly occupy the bands originated from carbon nanotubes.

\begin{figure}
\centerline{
\epsfxsize=3.0in \epsfbox{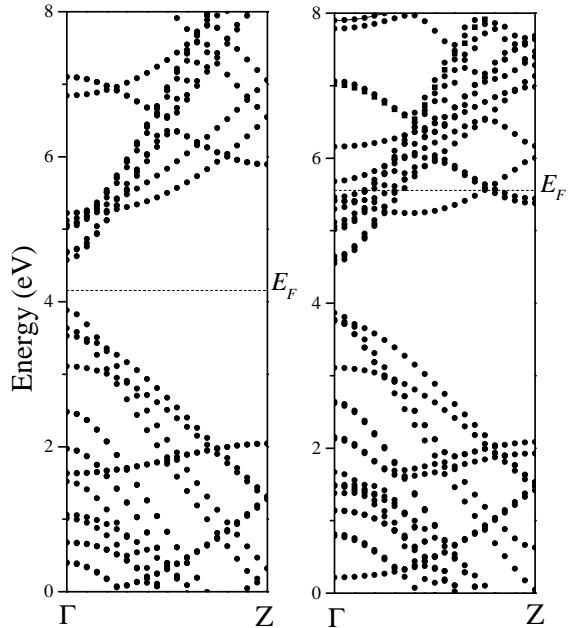}
}
\caption{Electronic band structures of pure (left) and Li intercalated (right)
(10,0) tube bundle. Most of the bands are not affected by Li atoms, whereas 
some new conduction bands are introduced. Most conduction electrons reside
on the bands associated with carbon nanotube, indicating charge transfer
from Li to nanotube.
}
\end{figure}

Our calculation on other nanotube bundles such as (10,10) tube bundle
(Li$_5$C$_{40}$) show similar charge transfer between Li ions and carbon
host. The observation of charge transfer agrees with previous
{\em ab initio} calculation on K doped individual small carbon nanotubes
\cite{11}. Similar effect is well known in alkali-metal-doped fullerens
\cite{18}. Experimentally, the charge transfer is supported by Raman
\cite{19} and NSR \cite{20} measurements on alkali-metal doped SWNT
materials. These suggest that the cohesion between Li and carbon nanotube is
mainly ionic. However, we have tried a simple model in which complete charge
transfer and uniform charge distribution on C atoms are assumed. Interactions
with different screening lengths were tested. We find that this simple model
is not sufficient to describe our results, indicating the importance of
screening and electron correlations.

To understand where the Li ions can be intercalated, we compare
the intercalation energy of two typical Li sites with high symmetry,
the center of tube and interstitial site of hexagonal lattice. The total
energy and equation of states of the tube bundle are calculated via 
Car-Parrinello electronic minimization method \cite{14}. The optimal distance 
between neighboring tubes are then determined. The intercalation energy
is obtained by subtracting the energy of the pure nanorope from the total
energy of intercalated system. SWNT bundles composed of (10,0), (8,8), (12,0) 
and (10,10) tubes are studied (Table I). In general, the energy of the Li
atoms inside the tube is found to be lower than or comparable to those outside 
the tube, implying that both the inside and outside of nanotube are favorable for Li
intercalation. For smaller tube, the center of tube
is less favorable because of the strong core repulse between Li ions and
carbon walls.

To study this issue further, we consider nanorope with different intercalation
density by putting certain number of lithium atoms at both interstitial
sites and those inside the tube. Typical Li configurations are briefly
illustrated in Table II. We find that the energies of both the Li sites
outside and inside the nanotube are comparable even up to rather
high intercalation density. For instance, the energy difference of nine Li
ions all inside or outside the (10,10) tube is only 0.36 eV per Li atom.
We also find that the intercalation energy is not sensitive to
Li arrangements at higher concentration. All these results imply that both
inside and outside of the tubes can be simultaneously intercalated to achieve
higher Li density.

\begin{table}
Table I. The energy $\Delta$E (which one Li per unit cell) between the Li
reside in interstitial site (a) or at the center of nanotube (c) 
(see Table II for coordinates). For smaller tube, the interstitial site 
is preferred while the center is better for larger tube.
\begin{center}
\begin{tabular}{ccccc}
Nanotube            & (10,0) & (12,0) & (8,8) & (10,10) \\ \hline
Radius (\AA)    & 3.91   & 4.70   & 5.42  & 6.78 \\
$\Delta$E (eV)  &  0.24  & -0.54   & -0.20  & -2.20 \\
\end{tabular}
\end{center}
\end{table}

In recent experiments, the intercalation density of as-prepared SWNTs bundles
sample was found as Li$_{1.6}$C$_6$ \cite{7}, and improved up to
Li$_{2.7}$C$_6$ after proper ballmilling \cite{8}. The experimental size of
carbon nanotubes is close to that of (10,10) tube. We suggest that the
ball-milling process creates defects or breaks the nanotube, allowing
the Li ions to intercalate inside of the tube. To understand the experimental
intercalation density, we study the intercalated (10,0), (12,0), (10,10) tube
bundles with intercalation density from 0 to 28 Li ions per unit cell.
Typical Li configurations are given in Table II. The intercalation
energy as function of intercalation density for nanoropes is compared
with that of graphite in Fig.4. We find that the intercalation energy per
carbon atom increases linearly with the intercalation density for
different tube bundles up to about Li$_{0.6}$C. In contrast, the Li
intercalation in graphite is already saturated at around Li$_{0.35}$C. We find
that the intercalation potentials defined by taking the derivative of 
intercalation energy with respect to intercalation density for all tube 
bundles are almost the same. It is comparable to the intercalation potential 
of graphite and about 0.1 eV higher than the formation energy of bulk lithium.

\newpage
\begin{table}
Table II. Examples of Li intercalation configurations in our study for (10,10)
tube bundle a $18.0\AA \times 18.0\AA \times 2.46\AA$ hexagonal unit cell.
For a given Li concentration, symmetric configuration is chosen to
maximize the Li-Li distance.
a-f are the site indexes and the number in the bracket is their degeneracy.
a,b,e sites are outside the tube and c,d,f are inside. x,y,z denote
fractional coordinates based on hexagonal lattice of nanorope. N is the
Li number in unit cell which contains 40 carbon atoms. X means there is a
Li atom on the site while O means no.

\begin{center}
\begin{tabular}{ccccccc}
Site &  a (2)  &  b (3) & c (1) & d (8) & e (6) & f(4) \\ \hline
 x   &   0.333 &   0.5  &   0   & 0.278 & 0.278 & 0.096 \\
 y   &   0.667 &   0.5  &   0   &   0   & 0.556 & 0.131 \\
 z   &    0.5  &   0    &   0   &   0   & 0.50  & 0.5   \\ \hline
N=2  &    X    &   O    &   O   &   O   &  O    &  O   \\
N=3  &    X    &   O    &   X   &   O   &  O    &  O   \\
N=5  &    X    &   X    &   O   &   O   &  O    &  O   \\
N=6  &    X    &   X    &   X   &   O   &  O    &  O   \\
N=13 &    X    &   X    &   O   &   X   &  O    &  O   \\
N=20 &    O    &   X    &   X   &   X   &  X    &  O   \\
N=24 &    X    &   X    &   X   &   X   &  X    &  X   \\
\end{tabular}
\end{center}
\end{table}

\begin{figure}
\centerline{
\epsfxsize=3.0in \epsfbox{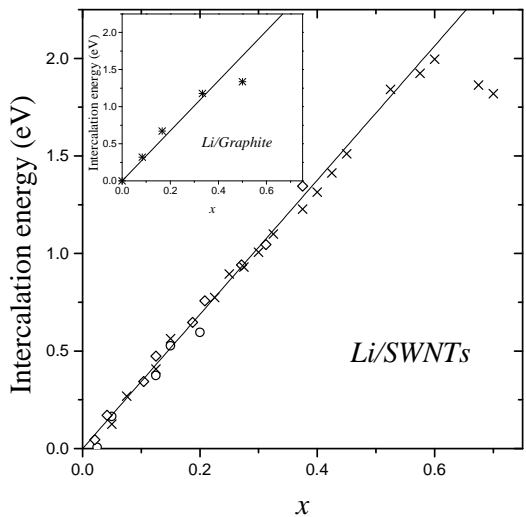}
}
\caption{Intercalation energy per carbon atom as a function of intercalation
density, $x$, of Li$_x$C systems. The insert figure show intercalation
energy for graphite. Crosses $-$ (10,10) tube bundle; open squares $-$
(12,0) tube bundle; open
circles $-$ (10,0) tube bundle. Solid lines are linear fits of the data
up to saturation density, Li$_{0.6}$C for nanorope and Li$_{0.35}$C for
graphite.
}
\end{figure}

From above results, we conclude that the nanorope has a higher capacity for
hosting the Li atoms if Li can penetrate into inside space of nanotube. This
agrees with the experimental finding that the
intercalation density in the ball-milled SWNT bundles can reach up
to Li$_{2.7}$C$_6$, much higher than LiC$_6$ in graphite \cite{7}.
The nature of higher Li capacity in nanotube can be related to the low carbon 
density in nanotube bundle. For example, the average atomic volume for carbon
in (10,10) tube bundle is about 60$\%$ larger than that of graphite.
The calculated saturation intercalation density is also about 60$\%$ higher
in (10,10) tube bundle than graphite. Additional understanding of high Li
concentration can be gained by examine the work function (WF) of nanotube.
Although we are unaware of any experimental measurements of WF on SWNT one
might expects that its WF to be close to that of C$_{60}$ thin films (4.85 eV)
\cite{21} and higher than WF of graphite (4.44 eV) \cite{22}.
Thus, the electron in nanorope has lower energy than those in graphite.

In summary, we have performed first principles calculations on the total energy
and electronic structures of Li intercalated SWNT nanoropes. The main conclusions
are: (1) almost complete charge transfer occurs between Li atoms and SWNTs;
(2) the deformation of nanotube structure after intercalation is relatively
small; (3) energetically inside of tube is as favorable as interstitial sites
for intercalation; (4) the intercalation potential of Li/SWNT is comparable to
the formation energy of bulk Li and independent of Li density up to
about Li$_{0.5}$C; (5) the intercalation density of SWNT bundle is
significant high than that of graphite. These results suggest that nanorope 
is a promising candidate material for anode in battery application. \\

This work is supported by the U.S. Army Research Office Grant
No. DAAG55-98-1-0298, the Office of Naval Research Grant No. N00014-98-1-0597
and NASA Ames Research Center. The authors thank Dr. O.Zhou and Mr. B.Gao for
helpful discussions. We gratefully acknowledge computational support from the
North Carolina Supercomputer Center.

\ \\
$^*$: zhaoj@physics.unc.edu \\
$^{\dagger}$: jpl@physics.unc.edu

\end{document}